\documentclass[epj]{webofc}
\usepackage[varg]{txfonts}   
\wocname{EPJ Web of Conferences}
\woctitle{CONF12}
\begin{document}
\selectlanguage{english} 
\title{Heavy Quark Coupled Channel Dynamics
  from Thermal Shifts}
%
%

\author{Enrique Ruiz
  Arriola\inst{1}\fnsep\thanks{\email{earriola@ugr.es}}~\footnote{Presenter
    at XIIth Conference on Quark Confinement and the Hadron Spectrum,
    August 29th to September 3rd 2016, Thessaloniki,
    Greece. E.R.A. thanks D.R. Entem for discussions.  This work is
    supported by Spanish Ministerio de Econom\'{\i}a y Competitividad
    and European FEDER funds under contracts FIS2014-59386-P and
    FPA2015-64041-C2-1-P, Junta de Andaluc\'{\i}a grant FQM-225, and
    Spanish Consolider Ingenio 2010 Programme CPAN
    (CSD2007-00042). The research of E.M. is supported by the European
    Union under a Marie Curie Intra-European fellowship
    (FP7-PEOPLE-2013-IEF) with project number PIEF-GA-2013-623006, and
    by the Universidad del Pa\'{\i}s Vasco UPV/EHU, Bilbao, Spain, as
    a Visiting Professor.} \and Lorenzo Luis Salcedo\inst{1}
\and Eugenio Meg\'{\i}as\inst{2,3}
}

\institute{Departamento de F\'{\i}sica At\'omica, Molecular y Nuclear and
  Instituto Carlos I de F\'{\i}sica Te\'orica y Computacional. Universidad
  de Granada, E-18071 Granada, Spain.
\and
  Max-Planck-Institut f\"ur Physik (Werner-Heisenberg-Institut), F\"ohringer Ring 6, D-80805 Munich, Germany
\and 
Departamento de F\'{\i}sica Te\'orica, Universidad del Pa\'{\i}s Vasco UPV/EHU, Apartado 644,  48080 Bilbao, Spain
}

\abstract{QCD at finite temperature below the phase transition should
  be determined in terms of colour singlet states such as hadrons and
  strings. We show how quark-hadron duality allows extracting sensible
  information concerning heavy quark and string breaking coupled
  channel dynamics from Polyakov loop correlators.}
\maketitle
\section{Introduction}
\label{intro}
A general principle of quantum statistical mechanics is the direct
relation between the thermodynamics of the system at equilibrium at
finite temperature $T$ and the spectrum of all possible states
$\Psi_n$ of a given Hamiltonian $H$, fulfilling $H\Psi_n = E_n
\Psi_n$~\cite{Huang:1987bk}. In a finite box with volume $V$
all states are discretized. Actually, the partition function counts every single state with a Boltzmann factor, 
\begin{equation}
Z= {\rm Tr} \, e^{-H/T} = \sum_n e^{-E_n/T}
\equiv e^{-F/T} \, , 
\end{equation}
where the Helmholtz free energy $F$ has been introduced.  As the
Hamitonian itself, $F(T,V)$ is determined up to an arbitrary constant
which may be fixed by a reference temperature, say $F_0= F(T_0,V)$.
All thermodynamical observables of the system may be deduced by the
standard relations for the entropy $S$ or the physically measurable
specific heat $c$ (which we rewrite as a fluctuation
~\cite{Huang:1987bk}),
\begin{equation}
S=-\partial_T F \, , \qquad c= T \partial_T S = \partial_T U = (\Delta
H)^2/T^2 >0 \, ,
\end{equation}
where $(\Delta H)^2 = \langle H^2 \rangle_T - \langle H \rangle_T^2 $.
In the thermodynamic limit, $V \to \infty$, we expect the volume to
factorize $F \to -V p$, $S \to V s$, $E = V \epsilon $. However,
when we add {\it test particles} $(Q_1, \dots, Q_N)$ to the system, the Hamiltonian and its eigenvalues are modified,  
$H \to H + \sum_{i=1}^NV_{Q_i}$, $E_n \to E_n + \Delta E_{n,Q}$, and
there appears a volume independent {\it shift} of the thermodynamic
quantities,
\begin{equation}
\Delta F \equiv F_{Q_1 \dots Q_N}  \, , \quad
\Delta S \equiv S_{Q_1 \dots Q_N} \, , \quad \Delta c \equiv c_{Q_1 \dots Q_N} \,.
\end{equation}

In this contribution we review how these thermodynamic shifts provide
information on heavy quark string breaking dynamics when a test 
heavy quark-antiquark pair separated at a fixed distance is
placed in the hot QCD vacuum below the phase
transition~\cite{Arriola:2014bfa,Arriola:2015gra,Megias:2015qya,Megias:2016bhk,Megias:2016onb}.
Note the these shifts {\it are not} true thermodynamic quantities. For
instance $c_Q = \Delta c$ is the difference of two positive
contributions but not necessarily positive itself~\footnote{A familiar
  example is given by the lowering of specific heat of snow in the
  presence of common salt (NaCl) which melts the mixture, meaning that
  the entropy shift  decreases with temperature. This effect is
  actually seen in QCD where $\Delta c \equiv c_Q < 0$ above the critical
  temperature~\cite{Megias:2016bhk}. Renormalization group equations
  for these quantities have been
  elaborated~\cite{Megias:2016onb}.}. 

\section{The hot QCD vacuum and the hadronic spectrum with u,d,s quarks}

\begin{figure}[h]
\centering
\hspace{-1cm}\includegraphics[width=3.5cm,height=4cm,clip]{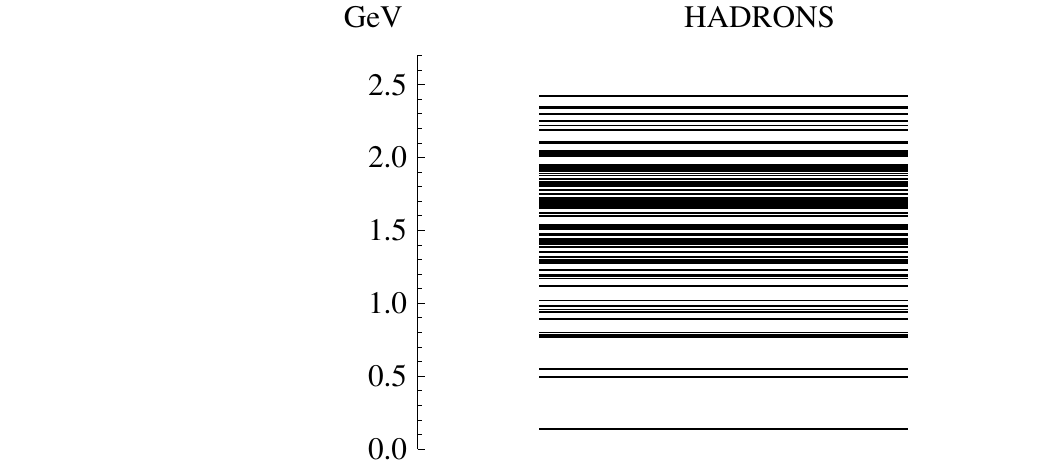}
\includegraphics[width=3.5cm,height=4cm,clip]{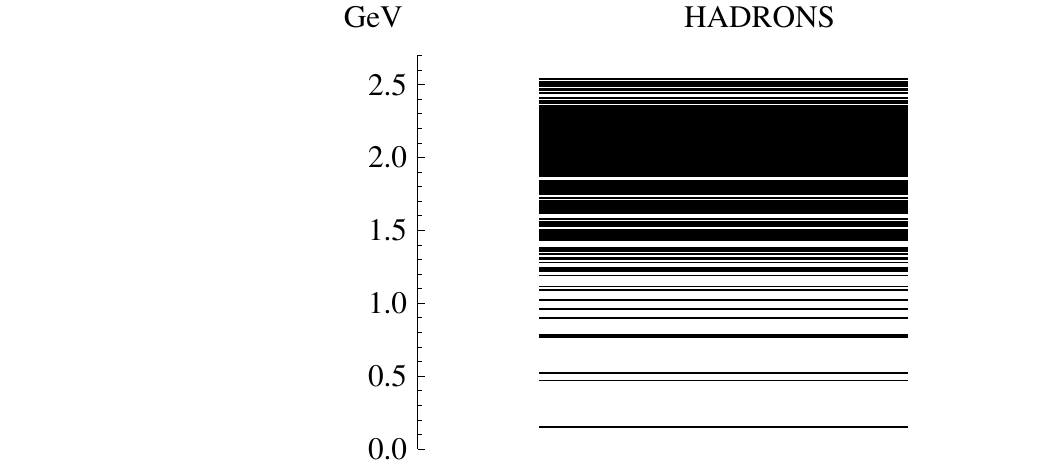}\hskip1cm
\includegraphics[width=6cm,height=4cm,clip]{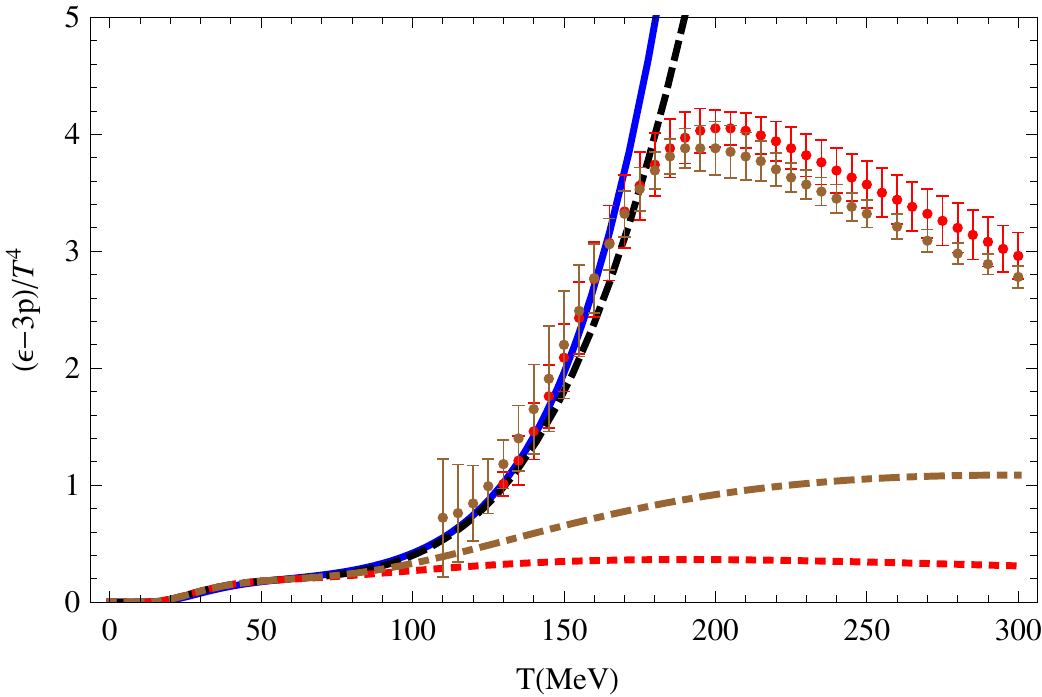}
\caption{Left panel: The PDG hadron spectrum~\cite{Olive:2016xmw}
  Middle panel: The RQM hadron
  spectrum~\cite{Godfrey:1985xj,Capstick:1986bm}.  Right panel: Trace
  Anomaly in lattice QCD~\cite{Borsanyi:2013bia,Bazavov:2014pvz} vs
  HRG using PDG (dashed) and RQM (solid) spectra. We also plot just
  the contribution of states with $M < 0.6 \, {\rm GeV}$ (dotted) and
  $M< 0.8 {\rm GeV}$(dotted-dashed).}
\label{fig-1}       
\end{figure}

Because of confinement the QCD {\it spectrum} is made of colour
singlet states. Up to electroweak corrections, we identify it with the
hadronic spectrum in the case of bound states. This identification
becomes a problematic issue in the case of a resonance where mass and
width depend on the particular production process and the
corresponding background definition in the experiment whereas in a
finite box such as in lattice QCD, a resonance is identified as a
volume independent energy shift~\footnote{It only becomes unambiguous
  as pole in the complex plane, what requires analytical continuation
  to the second Riemann sheet since complex energies are not directly
  measurable. This introduces a half-width ambiguity in the resonance
  mass~\cite{Arriola:2012vk}.}.  The question turns critical when the
number of states becomes large as it happens at finite
temperature. Particle states are termed hadrons when they are either
bound states or resonances and are phenomenologically listed in the
PDG~\cite{Olive:2016xmw}. So far, the current 2016 classification
echoes the naive quark model where mesons are $\bar q q $ and baryons
are $qqq$ states; the remaining ``further
states''~\cite{Olive:2016xmw} await consensual identification and will
be disregarded here. Therefore, it makes sense to compare with the
Relativized Quark Model (RQM) spectrum computed by Isgur, Godfrey and
Capstick in 1985~\cite{Godfrey:1985xj,Capstick:1986bm}.  We illustrate
a comparison between the 2016 PDG and the 1985 RQM in
Fig.~\ref{fig-1}.  As can be seen, both spectra are different, but it
will be shown that they are thermodynamically equivalent.

The thermodynamic approach provides a more global and balanced
comparison adapted to the physical situation such as
e.g. ultrarelativistic heavy-ions collisions where the equation of
state (EOS) plays a role in the hydrodynamics and kinetics of the
process~\cite{Florkowski:2010zz}.  From the QCD side it is known that
there exists a smooth crossover~\cite{Aoki:2006we} at about $T_c=150
{\rm MeV}$ (and zero chemical potential) which can be characterized by
a maximal violation of scale invariance of the trace anomaly ${\cal
  A}(T) = (\epsilon - 3 p)/T^4$ ( ${\cal A}(T)$ vanishes for massless
particles). The lattice calculations by the Bielefeld and Wuppertal
groups~\cite{Borsanyi:2013bia,Bazavov:2014pvz} comprise decades of
efforts which merging results are summarized in Fig.~\ref{fig-1} for
$120 {\rm MeV} \le T \le 300 {\rm MeV}$.

On the phenomenological hadronic side, one may assume that the EOS is
determined by hadron dynamics, where all resonance states which live
long enough contribute to the thermodynamics. This is unlike some
weakly bound states, such as e.g. the deuteron, which average out
within temperature intervals comparable to the binding
energy~\cite{Arriola:2015gra}. The Hadron Resonance Gas (HRG)
implements this idea as a multicomponent system of non-interacting
pointlike hadrons which is expected to work below $T_c$. If $M_n$ are
the masses of the hadrons one has
\begin{equation}
{\cal A}(T)= \frac{\epsilon-3 p}{T^4} = \sum_n g_n \int \frac{d^3 p}{(2\pi)^3}
\frac{E_n(p)- p \cdot \nabla_p E_n(p)}{e^{E_n(p)/T}+\eta_n} \,,
\end{equation}
where $g_n$ is the degeneracy, $E_n(p)= \sqrt{p^2+M_n^2}$ and
$\eta_n=\mp 1$ for Bosons and Fermions respectively. In
Fig.~\ref{fig-1} we compare the HRG when the spectrum is made out of
the states listed in the PDG~\cite{Olive:2016xmw} or taken from the
RQM~\cite{Godfrey:1985xj,Capstick:1986bm} with the recent lattice
calculations~~\cite{Borsanyi:2013bia,Bazavov:2014pvz}. This overall
agreement between the venerable RQM-based HRG, the upgraded PDG-based
HRG and the brand new QCD lattice calculations is truly impressive,
although not fully understood.

\section{Heavy quarks and quarkonia}

Heavy quark phenomenology is guided by abundant and varied information
on heavy quarkonia spectra and electroweak as well as strong
decays. Besides, their large mass allows to work in the
Born-Oppenheimer approximation and to treat their low-lying bound
states nonrelativistically. The Cornell potential works
phenomenologically well for hidden-- charm and bottom $\bar c c$ and
$\bar b b$ bound state mesons such as $\psi(nS)$ or $\Upsilon (nS)$
respectively. This potential has also been reproduced on the lattice
as an static energy accurately resembling that of a bosonic string of
length $r$ in the ground state~\cite{Necco:2001xg}
\begin{equation}
V_{\bar Q Q}(r)= \lim_{m_{\bar Q},m_Q \to \infty} \left[ E_{\bar Q Q}(r)- m_Q - m_{\bar Q} \right] = - \frac{\pi}{12 r} + \sigma r + c \,. 
\label{eq:cornell}
\end{equation}
The constant $c$ is ambiguous and depends on the renormalization
scheme. A renormalization prescription consists of imposing that the
$\Upsilon(1S)$ state wave function be node-less, in agreement with the
oscillation theorem (see
e.g.~\cite{Segovia:2011tb}). Phenomenologically it takes the common
value $c=-250 {\rm MeV}$ when $\sigma=(0.42 {\rm GeV})^2$ in the
RQM~\cite{Godfrey:1985xj,Capstick:1986bm} leading also to the $u,d,s$
spectrum of Fig.~\ref{fig-1} as well as charmonium and bottomonium
spectra.

Of course, some $\bar Q Q$ quarkonium states are unstable and undergo
a strong decay into heavy-light mesons $B=\bar q Q$ and anti-mesons
$\bar B = \bar Q q$ and can thus be interpreted as bound states in the
continuum, such as $\Upsilon (4 ^3S_1) \to \bar B B \equiv (\bar b u)
(\bar u b)$ (see Ref.~\cite{Ferretti:2013vua} for a RQM perspective
within the $^3P_0$ model ). The underlying mechanism is attributed to
string breaking between the colour conjugated sources generating
energetically favoured light $\bar q q$ pairs from the vacuum and
forming weakly interacting colour neutral states. In the
Born-Oppenheimer approximation this happens when the corresponding
energies cross $E_{\bar Q Q} (r_c) = M_{\bar q Q}+ M_{q \bar Q} $ {\it
  and} there is a transition potential $W (r) \equiv V_{\bar Q Q
  \leftrightarrow \bar B ;\bar B} (r)$.

\begin{figure}[h]
\sidecaption
\includegraphics[width=7cm,clip]{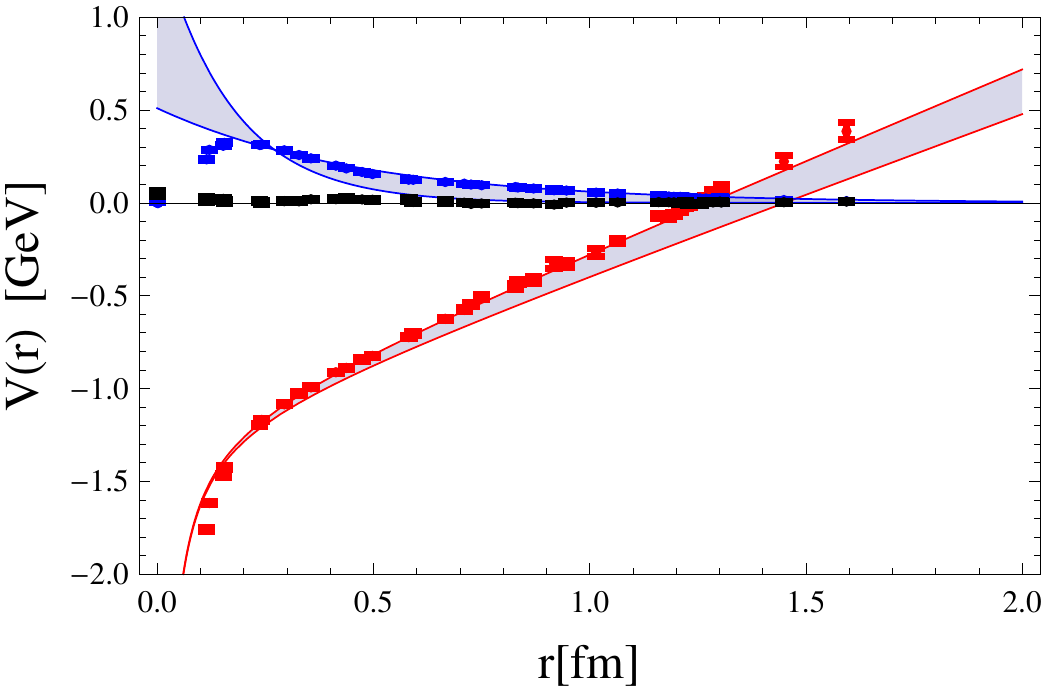}
\caption{The coupled channel potential with the diagonal matrix
  elements $V_{\bar QQ}(r)$ (red) and $V_{\bar B B}(r)$ (black) and
  the transition potential $V_{\bar Q Q \leftrightarrow \bar B B}(r)$
  (blue). We compare our two-mode extraction~\cite{Megias:2016onb}
  from finite temperature QCD free energies $F_{\bar Q Q} (r,T)$ with
  $N_f=2+1$ and physical pion mass~\cite{Borsanyi:2015yka} (gray
  bands), with the zero temperature determination and $m_\pi=700 {\rm
    MeV}$~\cite{Bali:2005fu} (error bars).  To ease the comparison we
  shift the lowest energies at large distances to zero.  }
\label{fig-2}       
\end{figure}

Quite generally this yields a multiple avoided crossing pattern, among
$B_n = (\bar q Q)_n$ and $\bar B_m = (\bar Q q)_m$ states with masses
$M_{B_n}= \Delta_n + m_Q $ and transition potentials
$W_{nm} (r) \equiv V_{\bar Q Q \leftrightarrow \bar B_n B_m} (r)$ and which can be 
described by a coupled channel static Hamiltonian 
\begin{equation}
{\cal H}(r) = 
\left(
\begin{array}{ccc}
  V_{\bar Q Q}(r)  &  W(r) &  \dots  \\
W(r)  &  2\Delta & \dots 
\\
\vdots  & \vdots & \ddots \\ 
\end{array} 
\right) \, 
\label{eq:V2states}
\end{equation}
and its corresponding eigenvalues $E_n^{\bar Q Q}(r)$. Some time ago
Drummond suggested a string breaking and mixing scenarios on the
lattice~\cite{Drummond:1998ar,Drummond:1998ir,Drummond:1998he} where
he proposed $W(r)= g e^{-m r}$ in the strong coupling limit. String
breaking has been observed in lattice QCD by the SESAM
Collaboration~\cite{Bali:2005fu} for pion masses of $m_\pi = 700 {\rm
  MeV}$ and using a linear combination of two modes $Q\bar Q$ and
$B\bar B$ confirming a sharp avoided crossing pattern (see
Fig.~\ref{fig-2}) at about $1.2 {\rm fm}$. In Fig.~\ref{fig-2} we show
the SESAM matrix elements of the coupled channel Hamiltonian of
Eq.~(\ref{eq:V2states}), where an almost constant $V_{\bar B B}(r) $
can be seen.


\section{Free energy and String breaking below the phase transition}

\begin{figure}[t]
\begin{center}
  \includegraphics[width=6cm,,height=4cm,clip]{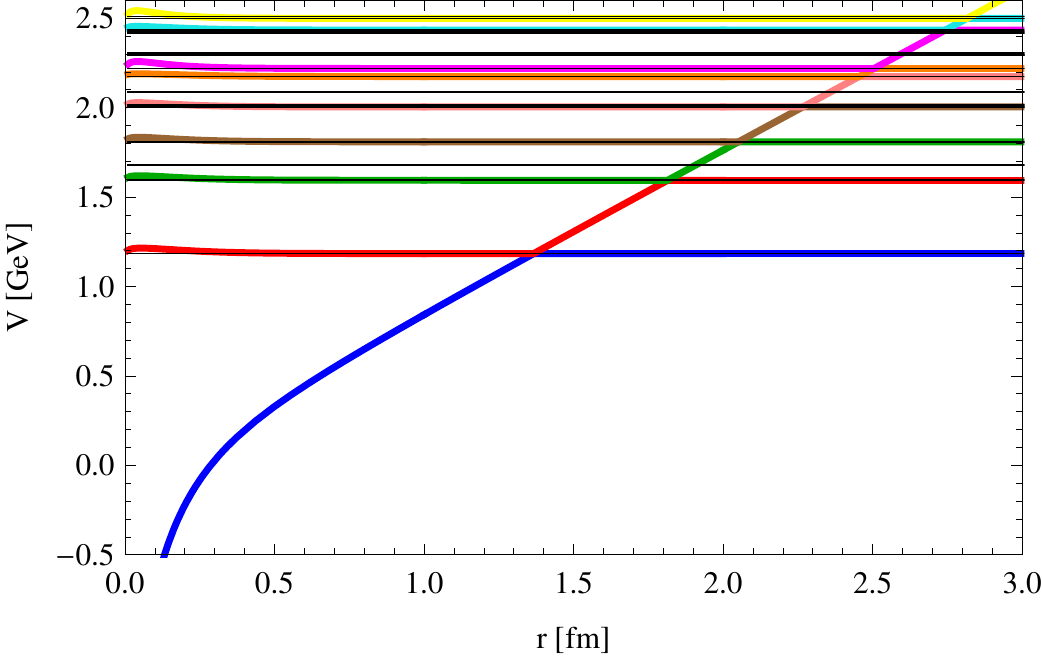}
  \includegraphics[width=6cm,,height=4cm,clip]{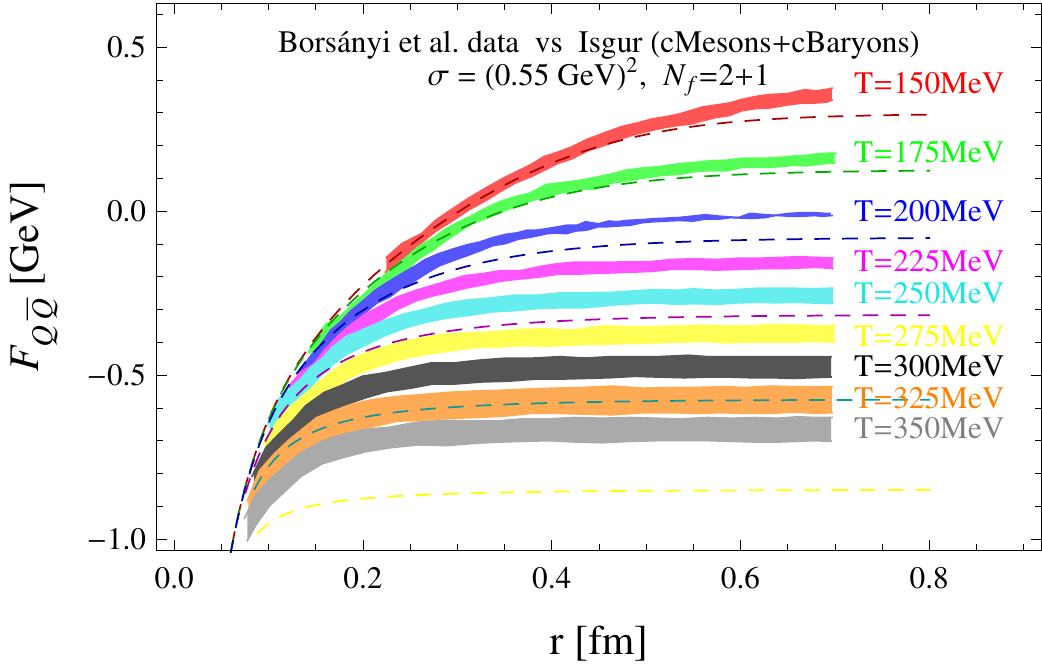}
\end{center}
\caption{Avoided level crossing spectrum with multiple mixing as a
  function of distance with RQM using c-quark (left pane) extracted
  in~\cite{Megias:2016onb} from the $\bar Q Q$ free energies obtained
  on the lattice~\cite{Borsanyi:2015yka} (right panel).}
\label{fig:plotcrossings}
\end{figure}

Relativistic heavy ions collisions have sufficient energy to produce
heavy quarks in the fireball made of light $u,d,s$ quarks and gluons.
On a more theoretical level, when we place two heavy colour conjugated
sources, say $\bar Q$ and $Q$ at a fixed distance $r$ in the hot
medium, we expect a thermodynamic shift of the equation of state, say
$F_{\bar Q Q} (r,T)$, which is given in terms of a Polyakov
loop correlator in the hot medium. In the static gauge we have
$\Omega(\vec r) = e^{igA_0 (\vec r) /T }$. The following  dual
relations hold~\cite{Megias:2016onb}
\begin{equation}
e^{-F_{\bar Q Q} (r,T)} = \langle {\rm tr} \Omega (r) {\rm tr}
\Omega (0)\rangle_T = \frac{Z_{Q \otimes\bar{Q}}(r,T)}{Z_0(T)} =
\frac{\sum_n e^{-E_n^{ Q \otimes\bar{Q}}(r)/T}}{\sum_n e^{-E_n^0/T}} =
\sum_n |\langle n,T | {\rm tr} \Omega |0,T\rangle|^2 \, e^{-r
  w_n(T)} \, ,
\end{equation}
where the last equality is a Kallen-Lehmann--type representation which implies
the new inequalities
\begin{equation}
\partial_r F_{\bar Q Q}(r,T) \ge 0 \,,\qquad \partial_r^2 F_{\bar Q Q}(r,T) \le 0 \,.
\end{equation}
The asymptotics depends on the distance $r$ vs the thermal wavelength
$1/T$. For $r T \ll 1$ the hot medium becomes irrelevant, $F_{\bar Q
  Q}(r,T) \to V_{\bar Q Q}(r)$, whereas for $r T \gg 1 $ cluster
decomposition implies $F_{\bar Q Q } (r , T) \to F_Q (T) + F_{\bar Q}
(T) $.
Here $F_Q (T) = F_{\bar Q} (T)$ is the single quark
free energy related to the expectation value of the Polyakov loop
which becomes a partition function at low
temperatures~\cite{Megias:2012kb}
\begin{equation} 
\langle {\rm tr} \Omega \rangle_T\equiv e^{-F_{Q} (T)/T} = \frac{Z_Q}{Z_0}
\approx  \sum_n e^{-\Delta_{n}/T} \,.
\end{equation}
Our normalization is such that $\langle {\rm tr}\Omega \rangle_T \to N_c$ for $T\to \infty$.  Below the phase transition the eigenvalues of
the $\bar Q q$ Hamiltonian can be saturated with heavy-light mesonic
states. Unfortunately, the number of such PDG states is too
small~\cite{Olive:2016xmw} and we use as in \cite{Megias:2012kb} the
corresponding RQM
states~\cite{Godfrey:1985xj,Capstick:1986bm}. Likewise, the $\bar Q Q$
system can be saturated with the static Hamiltonian sketched in
Eq.~(\ref{eq:V2states})
\begin{equation}
e^{-F_{Q \bar Q} (r,T)/T} \approx {\rm Tr} \, e^{-{\cal H} (r)/T}  \,.
\end{equation}
In Fig.~\ref{fig-2} we compare our results for the two-mode
Hamiltonian. From the heavy quark-antiquark free energy we extract a
string breaking profile,$V_{\bar Q Q \leftrightarrow \bar B B}(r)= W(r) = g e^{-m
  r}$, conforming to Drummond's
ideas~\cite{Drummond:1998ar,Drummond:1998ir,Drummond:1998he}. A fit of
the parameters of Eq.~(\ref{eq:cornell}), with $c=0$ and $\Delta=0.472
\, {\rm GeV}$ gives 
\begin{equation}
\sqrt{\sigma} = 0.424(14) {\rm GeV} \, ,  
\quad g=0.98(47) {\rm GeV} \, , \quad 
 m=0.80(38) {\rm GeV} \, , 
 \end{equation}
with $r_{\rm corr}(g,m)=0.98$. A complete avoided crossing pattern
based on inclusion of more states with the same quantum numbers as
$\bar QQ$, $V_{\bar Q Q \leftrightarrow \bar B_n B_m}(r)$, only
modifies the coupling but not the radial dependence and is shown in
Fig.~\ref{fig:plotcrossings}.  We note that the lowest branch displays
a more abrupt transition than assumed in (quenched) quark
models~\cite{Segovia:2011tb} but similar to the generalized screened
potential model~\cite{Gonzalez:2014nka,Gonzalez:2015hqa}.

\section{Effective elementarity and coarse graining}

Within this mixing scenario both the heavy quarks as well as the
heavy-light mesons $B$ and $\bar B$ are treated as pointlike, which
they are not. The range of annihilation generated by the profile
$W(r)=g e^{-m r}$, is about $r_s=1/m=0.25 {\rm fm}$, presumably
smaller than the size of heavy-light mesons.  This raises the question
of what is the meaning of the interaction when they overlap. In order
to elaborate on this, we will assume some underlying
light-quark--string dynamics. Taking massless light quarks,
$m_q=m_{\bar q}=0$, for simplicity we use the standard Salpeter
Hamiltonian for the charge conjugated heavy-light meson-antimeson
pairs $ H_{q \bar Q}= p_q + \sigma r_q $ and $ H_{\bar q Q}= p_{\bar
  q} + \sigma r_{\bar q} $. Thus, the spectrum is determined from the
equation $ ( p + \sigma r ) \Psi_n (r)= \Delta_n \Psi_n (r) $ whence
the virial theorem relation yields $\Delta_n = 2 \sigma \langle r
\rangle_n $~\cite{Arriola:2014bfa}. This has the interesting
interpretation that the string breaking happens at $ \sigma r_c =
\Delta_{\bar q Q} + \Delta_{\bar Q q} $ corresponding to $ r_c= 2
(\langle r_q \rangle + \langle r_{\bar q} \rangle ) $, i.e. when the
two extended mesons are about {\it twice} the overlapping
distance. Therefore, $B$ and $\bar B$, are well separated and the
overall separating distance $r$ is always well defined before they
annihilate into the $\bar Q Q$ pair.  This means that for $r > r_c$
the $\bar Q Q-\bar B B$ coupled channel dynamics can be treated as if
all intervening particles were elementary.

We turn to the shape of the string breaking profile function.  The
transition model we have in mind from the $\bar QQ$ to the $B\bar B$
is given by a contact interaction close in spirit to the $^3P_0$-model
$ v_{0 \to q\bar q} (\vec r_q - \vec r_{\bar q}) = g \delta^{(3)}
(\vec r_q - \vec r_{\bar q}) $. In reality, this function will be
smeared and, to see how, let us work in perturbation theory, where
$\Psi_{n,m} (r_q, r_{\bar q}) = \Psi_{B,n} (r_{\bar q}) \Psi_{\bar
  B,m} (r_{q}) $ and $\Delta_{nm}= \Delta_n + \Delta_m$. Then the
transition matrix element for two mesons located at a relative distance $r$ is
given by
\begin{equation}
W_{n,m}(r)= \int d^3 r_q  d^3 r_{\bar q} \Psi_{B,n} (r_{\bar q}+r/2) \Psi_{\bar B,m} (r_{q}-r/2)
v_{0 \to q\bar q} (r_q - r_{\bar q}) \,.
\end{equation}
This folding structure yields the profile function in momentum space, 
\begin{equation}
W (p) = \Psi_0^2(p) v_{0 \to \bar q q }(p) = W(0) \frac{m^4}{(m^2+p^2)^2}  \,.
\end{equation}
The very short range of the transition potential together with the
effective elementarity of the heavy-light mesons suggests a coarse
graining in the interaction, provided the typical wavelengths are longer
than $r_s = 1/m$. That this is the case can be seen in the nodal
structure of $\bar Q Q$ quark model wave functions up to
$n=5$~\cite{Segovia:2011tb}. This is very similar to the pattern
described by van Beveren and
Rupp~\cite{vanBeveren:2008rs} where they take $W(r)= G \delta (r-r_s)$
and for which they find $r_s \sim 0.4 {\rm fm}$, implying
a longer high momentum tail since in their case $W (p)= W(0) \sin (p
r_s)/(p r_s)$.  While the mass $m$ appearing in the transition
potential was identified with pion exchange in the lattice
calculation~\cite{Bali:2005fu} for pion masses of $m_\pi = 700 \, {\rm MeV}$, our results, extracted for physical $m_\pi$ free
energies~\cite{Borsanyi:2015yka} and phenomenology, suggest that this
is not the case. A more complete analysis will be presented elsewhere.


\end{document}